\documentclass[a4paper,11pt]{article}
\pdfoutput=1 

\usepackage{jcappub} 

\usepackage[T1]{fontenc} 

\usepackage{graphicx}
\usepackage{color}
\usepackage{epsfig,multicol,bbm} 
\usepackage{bm}
\usepackage{hyperref}

\title{Casimir dark energy, stabilization of the extra dimensions and Gauss-Bonnet term}

\author[a]{Pitayuth Wongjun}

\affiliation[a]{The Institute for Fundamental Study, Naresuan University, \\
Phitsanulok 65000, Thailand }
\emailAdd{pitbaa$@$gmail.com}

\abstract{Casimir dark energy model in five-dimensional and six-dimensional spacetime including non-relativistic matter and Gauss-Bonnet term is investigated. The Casimir energy can play the role of dark energy to drive the late-time acceleration of the universe while the radius of the extra dimensions can be stabilized. The qualitative analysis in radion picture in four-dimensional spacetime shows that the contribution from Gauss-Bonnet term will effectively slow down the radion field at the beginning time. Therefore, the radion field does not pass minimum point of the effective potential before the minimum of the potential exists. This leads to the stabilizing mechanism of the extra dimensions eventually.
 }

\begin{document}
\maketitle
\flushbottom

\section{Introduction}
The late-time acceleration of the universe is discovered by observing the behavior of the Supernovae type Ia (SN Ia) \cite{Riess:1998cb,Perlmutter:1998np}. Recent observations imply that about 72\% of the energy density of the universe consists of an unknown constituent called ``dark energy" \cite{Spergel:2003cb,Eisenstein:2005su,Ade:2013zuv}. One of the most useful and simple candidates of dark energy is cosmological constant. The cosmological constant can arise from a vacuum energy in particle physics theory and the results of this model can properly fit with the observational data. However, the energy scale of the vacuum energy calculated from particle physics theory is enormously larger than the observed value of the cosmological constant \cite{Weinberg:1989}.
It also encounters with coincidence problem, since the energy densities of cosmological constant and dark matter are significantly
different throughout the history of the universe while their energy densities are of the same order at the present
time \cite{Sahni:2008zza,Zeldovich:1967gd}. Therefore various kinds of dynamical model for dark energy are proposed in order to
explain the late-time acceleration of the universe \cite{Copeland:2006wr}.

A fundamentally theoretical framework that may be able to provide a description of the late-time acceleration of the universe is offered by string theory. Generally, string theory requires the present of extra dimensions. However, from the observation point of view, we live in four-dimensional spacetime. This leads to the fact that the extra dimensions have to be compactified. It is not easy to obtain the mechanism for stabilizing the extra dimensions while providing the viable model of dark energy \cite{Brandenberger:2005fb,Kodama:2005cz,Kodama:2006ay} and the lack of this mechanism is often called ``moduli stabilization problem". However, the recent search of this mechanism is still going on, for example in \cite{Shiu:2011zt, Flachi:2012rc}.

One of the most promising candidates of dark energy model that provides a solution for moduli stabilization problem is Casimir dark energy model \cite{Ponton:2001hq,Greene:2007xu,Burikham:2008fg}. Casimir energy is a vacuum energy emerging from imposing the boundary conditions to the quantum fluctuation fields. It is natural to interpret Casimir energy as dark energy instead of cosmological constant since Casimir energy can naturally emerge from the compactification mechanism. Moreover, this candidate of dark energy also provide the mechanism for stabilizing the extra dimensions automatically. However, in order to compare the results of the model to the standard history of the universe, we need to include non-relativistic matter content into the model. Unfortunately, adding the non-relativistic matter into the model will destroy the stabilizing mechanism of the extra dimensions \cite{Greene:2007xu,Chatrabhuti:2009ew}. The qualitative analysis shows that the minimum of the effective potential for the moduli field will disappear and its slope will increase. Hence, the moduli or radion field will roll down rapidly and then passes away from the minimum point before the minimum of the potential exists. This leads to the destabilization of the extra dimensions eventually. In order to restore the stabilizing mechanism, the modified Casimir dark energy model is investigated by adding the aether field into the model  \cite{Chatrabhuti:2009ew}. The effects of the aether field in the higher-dimensional spacetime are investigated in \cite{Rizzo:2005um,Carroll:2008pk} and one of the results in four-dimensional spacetime is that it will decrease the slope of the effective potential. In other words, it reduces the force acting on the radion field during matter-dominated period. Thus the radion field slowly rolls down at the beginning time and it has enough time for waiting the existence of the potential minimum at the late-time. This leads to the stabilization of the extra dimensions eventually. Unfortunately, the form of aether field which can provide this viable model is not stable \cite{Carroll:2009em,Himmetoglu:2008hx,Himmetoglu:2008zp,Himmetoglu:2009qi}. It is important to note that there is another class of study in stabilizing mechanism of the extra dimensions from Casimir energy \cite{Williams:2012au,Burgess:2012pc, Salvio:2012uu}. The Casimir energy from this model can also play the role of cosmological constant the drive the late-time acceleration of the universe.

In this paper, we will seek for another way to restore the stabilizing mechanism by considering the modification of gravity instead of adding an exotic matter field. For the modified gravity, we will consider the generalization of Einstein gravity namely ``Lovelock gravity" \cite{Lanczos:1938sf,Lovelock:1971yv,Zumino:1985dp}. Lovelock gravity is a generalization of Einstein gravity in higher-dimensional spacetime while keeping the full Einstein gravity with cosmological constant in four-dimensional spacetime. One of important properties of this modified gravity is that it still provides the second order derivative of the equations of motion and satisfies the conservation equation of matter field or, in other words, satisfies of the modified Bianchi identity. In five and six-dimensional spacetime, Lovelock gravity theory is reduced to Einstein-Gauss-Bonnet (EGB) gravity theory which is Einstein gravity theory including Gauss-Bonnet (GB) term. In fact, GB term can arise from string theory \cite{Antoniadis:1993jc,Campbell:1990fu}.  Therefore, it is worthy to investigate the effect of GB term on the stabilization of the extra dimensions in Casimir dark energy model and this is the aim of this work. We find the equations of motion in five-dimensional spacetime and then use the numerical computation to show that the extra dimension can be stabilized. The effective four-dimensional theory is obtained by Kaluza-Klein reduction \cite{Kaluza:1921tu,Klein:1926tv}. By using this result, we found that the contribution from GB term will effectively slows down the radion field at the beginning time and then the radion field does not pass minimum point before it exists eventually. We also investigate this mechanism in six-dimensional spacetime and found that the radius of the extra dimensions can be stabilized in the same manner as five-dimensional analysis.

The paper is organized as follows. We start with review of the Casimir dark energy model Section \ref{sec:Casimir}. In this section, we begin with introducing dark energy and then discuss the Casimir energy in five-dimensional spacetime. We close this section by showing that the extra dimension can be stabilized when non-relativistic matter is not included and will be destabilized when the non-relativistic matter is taken into account. In Section \ref{sec:EGB theory}, we review the Lovelock gravity theory in (4+n)-dimensional spacetime and specialize in the case of EGB gravity theory. The Kaluza-Klein reduction of EGB gravity theory is also reviewed in this section. In Section \ref{sec:CDE with GB}, we use the results of two previous sections to modify the Casimir dark energy model by including the GB term and show how the GB term effect the dynamics of radion field in both five and six-dimensional spacetime. Finally, we conclude the results in Section \ref{sec:conclusions}.

\section{Casimir dark energy model\label{sec:Casimir}}
According to observations, it is found that the universe is accelerating at late-time period \cite{Riess:1998cb,Perlmutter:1998np, Spergel:2003cb, Eisenstein:2005su, Ade:2013zuv}. Many theoretical models are proposed in order to describe this phenomena of the universe. In this section, we will review one of the theoretical models, the so called ``Casimir dark energy model" by following \cite{Greene:2007xu, Chatrabhuti:2009ew}. We will begin with the basic idea of dark energy model and then introduce Casimir energy which emerges from compactification of the extra dimension. We will show that such energy can drive the late-time accelerating universe and leads to the mechanism for stabilizing the extra dimension. However, when we include non-relativistic matter into the model, it is found that the stabilization mechanism will be destroyed.
\subsection{Late-time accelerating universe}

From homogenous and isotropic universe in the large scale, we adopt the flat Friedmann-Lema\^\i tre-Robertson-Walker (FLRW) metric in order to describe the dynamics of the universe,
\begin{equation}
ds^{2}=-dt^{2}+a^{2}(t)d {\bf x}^{2},\label{flrw}
\end{equation}
where $a(t)$ is a scale factor. The energy momentum tensors of all constitutes in the universe are assumed to be perfect fluid and can be written in the form
\begin{equation}
T^{\mu}_{\,\,\nu}=\text{Diag}(-\rho, p, p, p),
\end{equation}
where $\rho$ is the energy density of the fluid and $p$ the pressure. The Einstein field equation gives two equations
\begin{eqnarray}
H^{2}\equiv\Big(\frac{\dot{a}}{a}\Big)^{2}=\frac{1}{3M_{Pl}^2}\sum_{i}\rho_i,\label{G00}\\
\dot{H}=-\frac{1}{2 M_{Pl}^2}\sum_{i}(\rho_i+ p_i),\label{Gii}
\end{eqnarray}
where the summation is summed over all species constitutes in the universe and $H$ is the Hubble parameter. The conservation of the energy momentum tensor for each species can be expressed as
\begin{equation}
\dot{\rho_i}+3H(\rho_i + p_i)=0.\label{con}
\end{equation}
Using (\ref{G00}) and (\ref{Gii}), we obtain the acceleration equation
\begin{equation}
\frac{\ddot{a}}{a}=-\frac{1}{6 M_{Pl}^2}\sum_{i}(\rho_i+3p_i)=-\frac{1}{6 M_{Pl}^2}\sum_{i}\rho_i(1+3w_i),
\end{equation}
where $w_i$ is the equation of state parameter for each species. Therefore, the accelerating universe for each species requires that
\begin{equation}
w_i < -1/3,\label{acc}
\end{equation}
where, $w_{\text{matter}} = 0$ and $w_{\text{radiation}} = 1/3$. We can see that ordinary matter and radiation we already known cannot drive the accelerating universe. Many models the so called ``dark energy models" are constructed in order to explain the late-time accelerating universe, for example, quintessence models \cite{Wetterich:1987fm, Ratra:1987rm}, k-essence models \cite{ArmendarizPicon:2000dh,ArmendarizPicon:2000ah,Chiba:1999ka}, Galilean models \cite{Nicolis:2008in} and their generalization \cite{Deffayet:2009wt,Deffayet:2009mn}, vector field models \cite{Kiselev:2004py,ArmendarizPicon:2004pm}, three-form field models \cite{Koivisto:2009sd,Germani:2009iq} and holographic dark energy models \cite{Li:2004rb}. Moreover, there are many modified gravity models constructed in order to explain this accelerating universe, for example, $f(R)$ gravity models \cite{Capozziello:2003tk,Carroll:2003wy}, $f(G)$ gravity models \cite{Nojiri:2005jg} and recently investigation massive gravity models \cite{deRham:2010ik,deRham:2010kj}. Among various dark energy models, there has a natural model motivated from fundamental theory such as string theory called ``Casimir dark energy model" \cite{Ponton:2001hq,Greene:2007xu}. We will focus on this model in the next subsection.

\subsection{Casimir energy and its interpretation of dark energy}

Casimir energy is a vacuum energy emerging from imposing boundaries to the quantum fluctuation field in small scale \cite{Casimir:1948dh}. This energy is seem to be a physical energy since Casimir force can be observed in terrestrial experiments \cite{Sparnaay:1958wg, Lamoreaux:1996wh}. In this subsection, we will review the mathematical calculation and physical description of Casimir energy from the compactification of the extra dimension. Then we will interpret the Casimir energy as dark energy in order to drive late-time accelerating universe. Consequently, it is found that this dark energy model provides the mechanism for stabilizing the extra dimension. However, we will show that this mechanism will be destroyed when non-relativistic matter is taken into account.

Generally, Casimir energy can be derived from any number of the extra dimensions. In this subsection we will consider an ansatz in which a extra dimension is compactified as a circle $S^1$ and 5-dimensional spacetime can be thought of a product space between 4-dimensional flat FLRW  spacetime and this circle space. In six-dimensional spacetime with product space between four-dimensional flat FLRW spacetime and a simple two-dimensional torus, Casimir energy can be easily derived by using analogous manner of the derivation in five-dimensional spacetime \cite{Ponton:2001hq,Greene:2007xu}. However, the calculation in the non-trivial two-dimensional torus, for example torus which characterized by both its volume and shape, will be more complicated since we need to use other complicated mathematical tools in order to derive \cite{Ponton:2001hq, Burikham:2008fg}. In this paper we will use the results derived in five-dimensional spacetime to obtain the analogous one in six-dimensional spacetime with a simple torus, the torus which is characterized by only its volume. Line element of this ansatz can be written as
\begin{eqnarray}
ds^2 = -dt^2 + a^{2}(t) d {\bf x}^{2} + b(t)^2 dy^2,
\label{5-D metric}
\end{eqnarray}
where $b(t)$ denotes the radius of the compact fifth direction. The coordinates on $S^1$ are $0 \leq y \leq 2\pi$. Considering a simple massive scalar field living in this spacetime, the equation of motion for this scalar field is the Klien-Gordon equation
\begin{eqnarray}
(\partial_A\partial^A - m^2)\phi=0,
\end{eqnarray}
where $m$ is a mass of the scalar field and the uppercase Latin indices, $A, B, C, ...$ are five spacetime indices running as $\{0, 1, 2, 3, 5\}$. Since the fifth direction of the spacetime is compactified in a circle, we can impose the periodic boundary condition of the scalar field $\phi(y=0) = \phi(y=2\pi)$. The wave number in the compact direction will be quantized and then dispersion relation of the scalar can be written as
\begin{eqnarray}
-k^\mu k_\mu = m^2+\frac{\tilde{n}^{2}}{b^2},
\label{normal_dp}
\end{eqnarray}
where, $\tilde{n}\in\mathbb{Z}$ is the momentum number in the compact direction. The vacuum energy of the scalar field can be written as
\begin{eqnarray}
\widehat{E}_{cas} &=& \frac{1}{2}\left(\frac{L}{2 \pi}\right)^{3} \int d^{3}k
\sum_{\tilde{n}}\sqrt{ k^2+ m^2+ \frac{\tilde{n}^2}{b^2}},
\end{eqnarray}
where $L^{3}$ is the spatial volume of non-compact spacetime. The integration of summation above seems to be diverse since $k$ run from $0$ to $\infty$.
However, we can regularize this integration by using the Chowla-Selberg zeta function \cite{Elizalde1}. We will not show the explicit calculation  for this regularization procedure. The detail calculation can be seen in \cite{Chatrabhuti:2009ew}. The result of the regularization is finite and then will be interpreted as Casimir energy \cite{Ponton:2001hq}. For the massless and massive scalar fields, the energy density of each components can be respectively written as
\begin{eqnarray}
\widehat{\rho}_{cas}^{massless}&=&\frac{\widehat{E}_{cas}}{L^3 2\pi b}
=\frac{\Gamma(-2s+1)}{\Gamma(-1/2)}2^{2s}b^{2s-1}\pi^{3s-1}\zeta(-2s+1),\\
\widehat{\rho}_{cas}^{massive} &=& -2(2\pi
b)^{2s-1}(mb)^{(1-2s)/2}\sum_{\tilde{n}=1}^{\infty}\tilde{n}^{(2s-1)/2}K_{(1-2s)/2}(2\pi
b m\tilde{n}),
\end{eqnarray}
where $\zeta$ is the zeta function, $\Gamma$ is the gamma function and $K_\nu(x)$ is the modified Bessel function. We also define new parameter for convenience $s= -2$. For other bosonic fields, it is found that Casimir energy can be written in the same form as scalar field. Moreover, the contribution from fermionic fields is also in the same expression with scalar field but it has a negative sign. In order to interpret Casimir energy as dark energy, we can expect the total Casimir energy density as a potential term of radion field in 4-dimensional spacetime. The radion field with the potential contributed from Casimir energy density can play the role of dark energy if there exists a positive minimum of the potential. In order to obtain the positive minimum of the potential, one has to choose the proper contribution from both massive/massless boson and fermion and the mass ratio between boson and fermion $\bar{\lambda} = m_b/m_f$.
Phenomenologically, we choose the contribution from each species as
\begin{equation}
\rho_{Cas} = 5 \rho_{boson}^{massless} + 8 \rho_{fermion}^{massless} + 8 \rho_{boson}^{massive} + 8 \rho_{fermion}^{massive}.
\label{sum_casimir}
\end{equation}
The number of the degrees of freedom of the massless boson comes from the graviton in 5-dimensional spacetime which has five degrees of freedom. For other species the number of the degrees of freedom are chosen in order to obtain the minimum of the potential.

\subsection{Dynamics of Casimir dark energy}
In order to obtain the dynamics of the Casimir dark energy, we add the energy momentum tensor contributed from the Casimir effect into the Einstein field equation. The general form of the Casimir energy momentum tensor which is compatible with the metric in the equation (\ref{5-D metric}) can be written as \cite{Greene:2007xu}
\begin{equation}
T^{\mu}_{~\nu(Cas)}= \text{diag}(- \rho_{Cas}, p_a, p_a, p_a, p_b,...,p_b )\label{Cas emt}
\end{equation}
where $p_a$ and $p_b$ are the Casimir pressure in the non-compacted and compacted dimension respectively. These pressures can be defined as \cite{Greene:2007xu}
\begin{eqnarray}
p_a &\equiv& -\frac{\partial}{\partial V_a}\Big(\rho_{Cas} V_a \Big),\\
p_b &\equiv& -\frac{\partial}{\partial V_b}\Big(\rho_{Cas} V_b\Big),
\end{eqnarray}
where $V_a \propto a^{d-n}$ and $V_b \propto b^{n}$. Here, $d$ is the number of all spatial dimensions and $n$ is the number of the extra dimensions and $d= 4, n= 1$ for this model. These definitions automatically yield the cosmological constant behavior in 4-dimensional spacetime while $p_a = -\rho_{Cas}$ and $p_b = -\rho_{Cas} - b\partial_b \rho_{Cas}$. The conservation equation of the energy momentum tensor reads
\begin{eqnarray}
\dot{\rho}_{Cas} +  3H_a(\rho_{Cas} + p_a) + n H_b(\rho_{Cas} + p_b) = 0,
\end{eqnarray}
where $H_a = \dot{a}/a$ and $ H_b = \dot{b}/b$. Substituting the energy momentum tensor into the Einstein field equation, one obtains
\begin{eqnarray}
3H_a^2+ \frac{n}{2}(n-1)H^2_b +3nH_a H_b &=& M^{-(n+2)}_{*}\rho_{Cas},\,\,\,\,\,\,\,\label{eom1nomatter}\\
n\frac{\ddot{b}}{b}+2\frac{\ddot{a}}{a}+\frac{n}{2}(n-1)H^2_b+H_a^2 +2nH_a H_b &=& -M^{-(n+2)}_{*}p_a,\label{eom2nomatter}\\
(n-1)\frac{\ddot{b}}{b}+3\frac{\ddot{a}}{a} +\frac{(3n-2)(n-1)}{2}H^2_b+ 3H_a^2 +3(n-1)H_aH_b &=& - M^{-(n+2)}_{*} p_b. \label{eom3nomatter}
\end{eqnarray}
where $M_{*}$ is the mass scale in (4+n)-dimensional spacetime. Note that we generalized Einstein field equation into (4+n)-dimensional spacetime for convenience.

\begin{figure}[htp]
\centering
\includegraphics[width=0.95\textwidth]{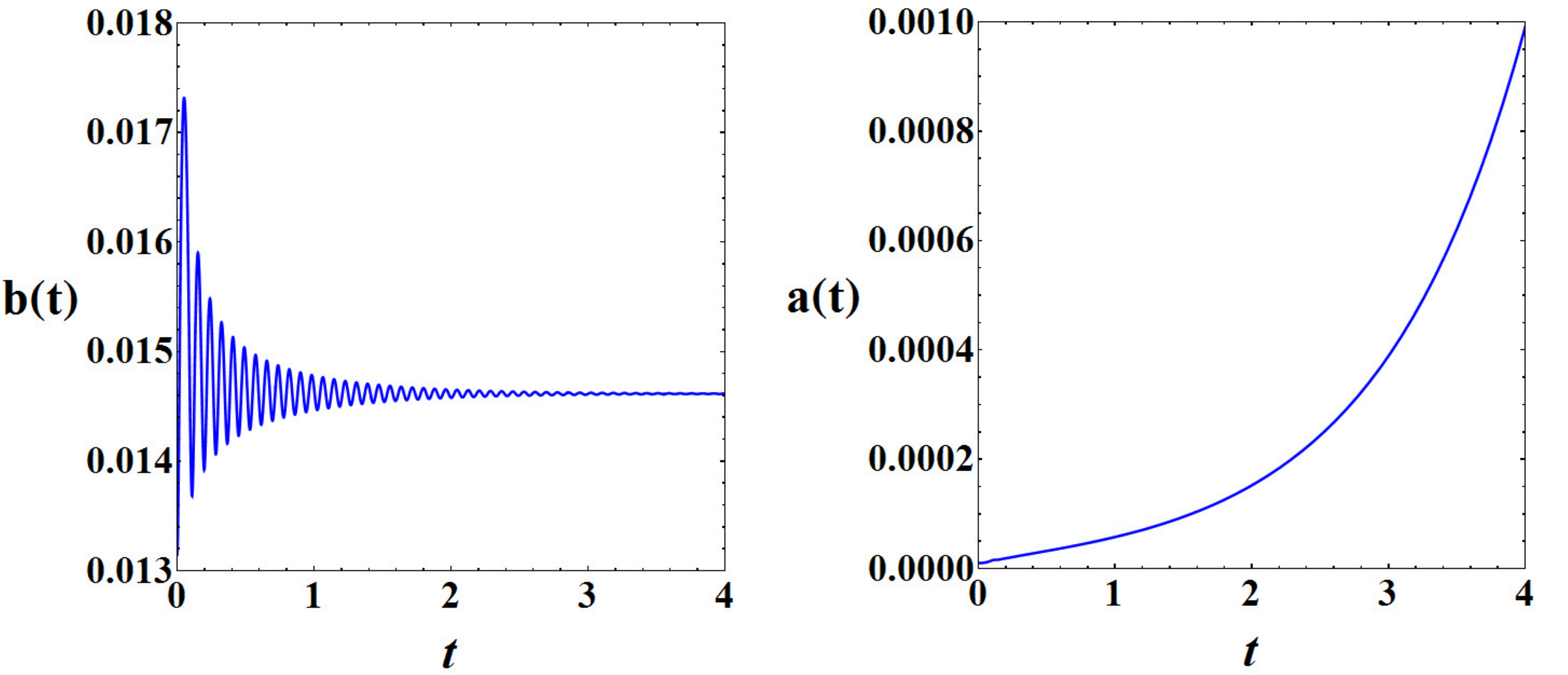}
\caption[The evolution of the radius of extra dimension and the scale factor without matter and Gauss-Bonnet term]{The evolution of radius of the extra dimension (in the left panel) and the scale factor (in the right panel) in the Casimir dark energy model without non-relativistic matter. From the left panel, the radius of the extra dimension can be stabilized and from the right panel our 3-spatial universe is accelerated. }\label{HbHa_Nomatter_NoGB}
\end{figure}

\begin{figure}[htp]
\centering
\includegraphics[width=0.95\textwidth]{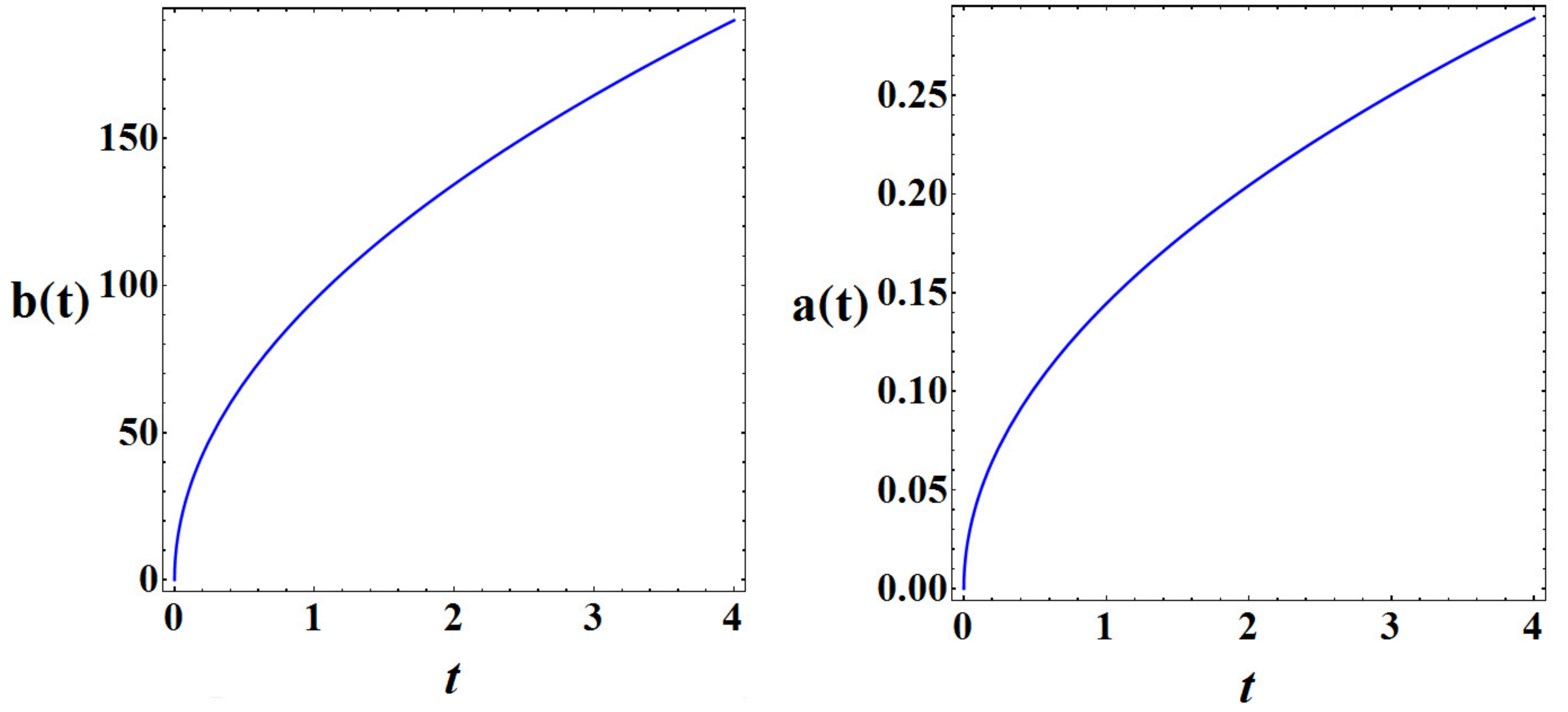}
\caption[The evolution of radius of the extra dimension and the scale factor without matter and Gauss-Bonnet term]{The evolution of radius of the extra dimension (in the left panel) and the scale factor (in the right panel) in the Casimir dark energy model including non-relativistic matter. From the left panel, the radius of the extra dimension cannot be stabilized and from the right panel our 3-spatial universe cannot be accelerated. }\label{HbHa_matter_NoGB}
\end{figure}

For $n=1$ the numerical results of these equations are shown in Figure \ref{HbHa_Nomatter_NoGB}. From this figure, we can see that the radius of the extra dimension can be stabilized at $b(t)\sim 0.0145$ and the scale factor is accelerated implying that our universe is accelerating where $\bar{\lambda}$ is set to be $\bar{\lambda} = 0.514$. To obtain the realistic cosmological history of the universe, we have to add the contribution from non-relativistic matter. The Einstein equations including non-relativistic matter in 5-dimensional spacetime can be written as
\begin{eqnarray}
3H_a^2+ 3H_a H_b &=& M^{-3}_{*}(\rho_{Cas} + \rho_m),\label{eom1matter}\\
\frac{\ddot{b}}{b}+2\frac{\ddot{a}}{a}+H_a^2 +2H_a H_b &=& -M^{-3}_{*}p_a,\label{eom2matter}\\
3\frac{\ddot{a}}{a}+ 3H_a^2 &=& - M^{-3}_{*} p_b,\label{eom3matter}
\end{eqnarray}
where $\rho_m $ is the energy density of non-relativistic matter in five-dimensional spacetime. The energy density of non-relativistic matter in (4+n)-dimensional spacetime can be written as
\begin{equation}
\rho_m = \left(\frac{b_{min}}{b}\right)^n\frac{\rho_{m0}}{a^3},
\end{equation}
where $\rho_{m0}$ is the energy density of non-relativistic matter nowadays corresponding to ($b=b_{min}$ and $a=1$). From the observational data, $\rho_{m0}^{(4)} = (2.8/7.2) \rho_{\Lambda} = (2.8/7.2) (2.3\times 10^{-3} eV)^{4}$. Therefore $\rho_{m0} = (2.8/7.2) \rho_{Cas}(b=b_{min})$, since $\rho_{\Lambda} = (2 \pi b_{min})^n \rho_{Cas}(b=b_{min})$ and $\rho_{m0}^{(4)} = (2 \pi b_{min})^n \rho_{m0}$. Using this relation the energy density of non-relativistic matter in (4+n)-dimensional spacetime can be written as
\begin{equation}
\rho_m = \frac{2.8}{7.2}\left(\frac{b_{min}}{b}\right)^n\rho_{Cas}(b=b_{min})a^{-3}.
\end{equation}
The numerical results of the evolution of $b(t)$ and $a(t)$ for equations (\ref{eom1matter})-(\ref{eom3matter}) are shown in Figure \ref{HbHa_matter_NoGB}. We can see that the radius of the extra dimension cannot be stabilized and the scale factor will not be accelerated.
The mechanism for destabilizing of the extra dimension will be examined by considering the effective potential of the radion field in four-dimensional spacetime. The minimum of the effective potential for the radion field does not exist at the early time since the contribution of non-relativistic matter is dominant. Therefore, the radion field will roll down and pass away from the minimum point before it exists \cite{Greene:2007xu}. We will consider this issue in detail in section \ref{sec:CDE with GB}. By including the effect of aether field, the stabilization of the extra dimension can be restored \cite{Chatrabhuti:2009ew}. However, the aether field by itself is not stable \cite{Carroll:2009em, Himmetoglu:2008hx, Himmetoglu:2008zp, Himmetoglu:2009qi}. Hence the stabilizing mechanism by including the aether field may not be trustable.

\section{Einstein-Gauss-Bonnet theory \label{sec:EGB theory}}
In this section, we briefly review the concept of Lovelock invariance. This leads to a generalization of Einstein's gravity theory by keeping second order equations of motion and covariant conservation of matter field. This generalization does not change Einstein's gravity theory in four-dimensional spacetime but gives a nontrivial modification of Einstein's gravity theory when the theory is considered in higher-dimensional spacetime. For five or six-dimensional spacetime, it is known that this generalization is Einstein-Gauss-Bonnet (EGB) theory. We will consider this theory especially in this section since we restrict our attention in Casimir dark energy model emerging from compactification of spacetime dimensions from five and six to four. The Kaluza-Klein compactification of EGB theory is also reviewed in the final part of this section.

\subsection{Lovelock invariance \label{subsec:Lovelock}}

General relativity is a gravity theory based on second order equation of motion called Einstein field equation, $G_{\mu\nu} = R_{\mu\nu} - 1/2 R g_{\mu\nu} = M^{-2}_{Pl} T_{\mu\nu}$, and satisfied the matter conservation equation, $\nabla_\mu T^{\mu\nu} = 0$ corresponding to Bianchi identity. The effective action for this gravity theory is Einstein-Hilbert action,
\begin{equation}
S_{EH} = \int d^4x \sqrt{-g} \frac{M^{2}_{Pl}}{2}R.
\end{equation}
Generally, higher order covariant scalars constructed from the metric tensor such as $R^2$, $ R_{\mu\nu} R^{\mu\nu}$ and  $R_{\mu\nu\rho\sigma}R^{\mu\nu\rho\sigma}$ will give higher derivative order of the equations of motion. However, there is a linear combination of the higher order covariant scalar which provides the second order equation of motion. In D-dimensional spacetime, the action of this linear combination can be written as
\begin{equation}
S_{L} = \int d^D x \sqrt{-g} \frac{M^{D-2}_{*}}{2} L_D,
\end{equation}
where $M_{*}$ is the fundamental mass scale of D-dimensional theory. $L_D$ is the Lovelock lagrangian in D-dimensional spacetime defined as
\begin{eqnarray}
L_D &\equiv& \sum_{0 \leq p < D/2 } \alpha_p \lambda^{2(p-1)}L_{(p)},\\
L_{(p)}&=& \frac{1}{2^{p}} \delta^{\mu_1 \mu_2 \mu_3 ...\mu_{2p}}_{\nu_1 \nu_2 \nu_3...\nu_{2p} } R^{\nu_1 \nu_2}_{\,\,\,\,\,\,\,\,\,\, \mu_1 \mu_2}...R^{\nu_{2p-1} \nu_{2p}}_{\,\,\,\,\,\,\,\,\,\,\,\,\,\,\,\,\,\,\mu_{2p-1}\mu_{2p}},
\end{eqnarray}
where $\alpha_p$ are dimensionless parameters, $\lambda$ is a length scale of parameter and $L_{(0)} = 1$.
We note that $\delta^{\mu_1 \mu_2 \mu_3 ...\mu_{2p}}_{\nu_1 \nu_2 \nu_3...\nu_{2p} }$ is the Kronecker symbol of order $2p$ defined as
\begin{eqnarray}
\delta^{\mu_1 \mu_2 \mu_3 ...\mu_{2p}}_{\nu_1 \nu_2 \nu_3...\nu_{2p} } = \det (\delta^\mu_\nu),
\end{eqnarray}
where $\mu$ and $\nu$ stand for $\mu_1,...\mu_{2p}$ and $\nu_1,...\nu_{2p}$ respectively, for example, $\delta^{\mu_1 \mu_2}_{\nu_1 \nu_2} = \delta^{\mu_1}_{\nu_1}\delta^{\mu_2}_{\nu_2}-\delta^{\mu_1}_{\nu_2}\delta^{\mu_2}_{\nu_1}$. For the zero order, it corresponds to cosmological constant with $\Lambda = \lambda^{-2}$ since $L_{(0)} = 1$. For the first order, we have
\begin{eqnarray}
L_{(1)}&=& \frac{1}{2} \delta^{\mu_1 \mu_2}_{\nu_1 \nu_2 } R^{\nu_1 \nu_2}_{\,\,\,\,\,\,\,\,\,\, \mu_1 \mu_2},\\ \nonumber
\,&=& \frac{1}{2}(\delta^{\mu_1}_{\nu_1}\delta^{\mu_2}_{\nu_2}-\delta^{\mu_1}_{\nu_2}\delta^{\mu_2}_{\nu_1})R^{\nu_1 \nu_2}_{\,\,\,\,\,\,\,\,\,\, \mu_1 \mu_2},\\ \nonumber
\,&=&\frac{1}{2}( R + R)= R.
\end{eqnarray}
Therefore, in four-dimensional spacetime, the most general action which provides the second order equation of motion is
\begin{eqnarray}
L_4 = \alpha_0 \Lambda + \alpha_1 R.
\end{eqnarray}
In order to obtain Einstein-Hilbert action with the cosmological constant, we can set $\alpha_0 = -2$ and $\alpha_1 = 1$ and if we will consider only pure Einstein gravity theory we can set $\alpha_0 = 0$ and $\alpha_1 = 1$.
For five or six-dimensional spacetime, we have to include the $L_{(2)}$ term into the action. Thus $L_{(2)}$ can be expressed as
\begin{eqnarray}
L_{(2)}&=& \frac{1}{4} \delta^{\mu_1 \mu_2 \mu_3 \mu_4}_{\nu_1 \nu_2 \nu_3 \nu_4} R^{\nu_1 \nu_2 }_{\,\,\,\,\,\,\,\,\,\, \mu_1 \mu_2}\, R^{\nu_3 \nu_4 }_{\,\,\,\,\,\,\,\,\,\, \mu_3 \mu_4},\\ \nonumber
\,&=& R^2 -4 R^{\nu_1 \nu_2 }R_{\nu_1 \nu_2 }+R^{\mu_1 \mu_2 }_{\,\,\,\,\,\,\,\,\,\, \mu_3 \mu_4}\, R^{\mu_3 \mu_4 }_{\,\,\,\,\,\,\,\,\,\, \mu_1 \mu_2}.
\end{eqnarray}
This Lagrangian density is called ``Gauss-Bonnet" Lagrangian or ``Lanczos" Lagrangian. Gravity theory that includes this Lagrangian into the Einstein-Hilbert action is called ``Einstein-Gauss-Bonnet" (EGB) theory. Note that we can add Gauss-Bonnet Lagrangian in to Einstein-Hilbert action in 4-dimensional spacetime but it does not contribute to the equation of motion.

In a similar way, the equation of motion in D-dimensional spacetime can be written in the compact formula as follow
\begin{eqnarray}
E^\mu_{\nu,D} &\equiv& \sum_{0 \leq p < D/2 } \alpha_p \lambda^{2(p-1)}E^\mu_{\nu,(p)},\\
E^\mu_{\nu,(p)}&=& -\frac{1}{2^{(p+1)}} \delta^{\mu\,\mu_1 \mu_2 \mu_3 ...\mu_{2p}}_{\nu\,\nu_1 \nu_2 \nu_3...\nu_{2p} } R^{\nu_1 \nu_2}_{\,\,\,\,\,\,\,\,\,\, \mu_1 \mu_2}...R^{\nu_{2p-1} \nu_{2p}}_{\,\,\,\,\,\,\,\,\,\,\,\,\,\,\,\,\,\,\mu_{2p-1}\mu_{2p}},
\end{eqnarray}
where $E^\mu_{\nu,(0)} = -\frac{1}{2} \delta^\mu_\nu$. The conservation of this tensor corresponding matter conservation equation is also obtained,
\begin{eqnarray}
\nabla_\mu E^\mu_{\nu,(p)}=0.
\end{eqnarray}
The explicit form of the first order Lanczos tensor can be written as
\begin{eqnarray}
E^\mu_{\nu,(1)} &=& -\frac{1}{4} \delta^{\mu\,\mu_1 \mu_2 }_{\nu\,\nu_1 \nu_2 } R^{\nu_1 \nu_2}_{\,\,\,\,\,\,\,\,\,\, \mu_1 \mu_2},\\ \nonumber
\,&=& R^\mu_\nu -\frac{1}{2}R \delta^\mu_\nu, \\ \nonumber
\,&=& G^\mu_\nu.
\end{eqnarray}
Equating this tensor to energy momentum tensor of matter, $ M^{-2}_{Pl} T^{\mu(m)}_{\nu}$, it recovers Einstein field equation. Similarly, if we take into account the zero order, we will obtain the Einstein field equation with cosmological constant. For the second order one, we have the Lanczos tensor as follow
\begin{eqnarray}
H^\mu_\nu=E^\mu_{\nu,(2)} &=& -\frac{1}{8} \delta^{\mu\,\mu_1 \mu_2 \mu_3 \mu_4 }_{\nu\,\nu_1 \nu_2 \nu_3 \nu_4 } R^{\nu_1 \nu_2 }_{\,\,\,\,\,\,\,\,\,\, \mu_1 \mu_2}\, R^{\nu_3 \nu_4 }_{\,\,\,\,\,\,\,\,\,\, \mu_3 \mu_4},\\ \nonumber
\,&=& 2\Big( R^{\mu \mu_1 \mu_2 \mu_3 }R_{\nu \mu_1 \mu_2 \mu_3 } - 2 R^{\mu_1 \mu_2 }R^\mu_{\,\,\mu_1\nu\ \mu_2} -2R^{\mu \mu_1 }R_{\nu \mu_1 } +R R^\mu_\nu \Big)-\frac{1}{2}L_{(2)}\delta^\mu_\nu.
\end{eqnarray}
Now we can summarize that the most general gravity theory in five or six-dimensional spacetime with keeping second order equations of motion and satisfying covariant conservation equation is Einstein-Gauss-Bonnet theory. The action of this theory including matter field can be expressed as
\begin{equation}
S_{EGB} = \int d^D x \sqrt{-g} \left( \frac{M^{D-2}_{*}}{2} (\alpha_0 \lambda^{-2} + \alpha_1 R + \alpha_2 \lambda^2 \mathcal{G}) + L_m \right),\label{D-action-EGB}
\end{equation}
where $\mathcal{G} = L_{(2)} = R^2 -4 R^{\nu_1 \nu_2 }R_{\nu_1 \nu_2 }+R^{\mu_1 \mu_2 }_{\,\,\,\,\,\,\,\,\,\, \mu_3 \mu_4}\, R^{\mu_3 \mu_4 }_{\,\,\,\,\,\,\,\,\,\, \mu_1 \mu_2}$ is Gauss-Bonnet term and $L_m$ is matter field Lagrangian. The equations of motion corresponding to this action is
\begin{equation}
-\frac{\alpha_0}{2} \lambda^{-2} \delta^\mu_\nu + \alpha_1 G^\mu_\nu + \alpha_2 \lambda^2 H^\mu_\nu = M^{2-D}_{*} T^{\mu(m)}_{\nu}, \label{D-eom-EGB}
\end{equation}
where $T_{\mu\nu}^{ (m)} = 2 \delta (\sqrt{-g}L_m)/\delta g^{\mu\nu}$ is the energy momentum tensor of matter field.
It is important to note that Lovelock invariance can be considered in terms of Vierbein or tetrad formalism. In this formalism, the Lovelock Lagrangian can be constructed from powers of the curvature two-form. The advantage points of this formalism are that it provides the clear geometric interpretation and it is easy to show that the equations of motion corresponding to Lovelock Lagrangian are second order.

\subsection{Kaluza-Klein compactification of EGB theory \label{subsec:KK of EGB theory}}
Since observations suggest that the universe is in four-dimensional spacetime, the higher-dimensional spacetime has to be compactified. In this subsection, we consider the Kaluza-Klein compactification from (4+n)-dimensional spacetime to 4-dimensional spacetime where $n$ is a number of the extra dimensions. We also restrict our attention only in diagonal metric of the internal extra dimensions for simplicity. The metric can be written as
\begin{eqnarray}
ds^2 = g_{AB} dx^A dx^B = e^{2 \alpha \phi}\bar{g}_{\mu\nu} dx^\mu dx^\nu + e^{2 \beta \phi} \tilde{g}_{ab} dy^a dy^b,
\label{4+n metric}
\end{eqnarray}
where the indices $A, B, ..$ run over all D-dimensional spacetime, the indices $\mu, \nu, ...$ run over the $(3+1)$-dimensional spacetime and indices $a, b, ...$ run over the internal space $n$ dimensions. For simplicity, we assume that $\bar{g}_{\mu\nu}$ and $\tilde{g}_{ab}$ are diagonal and $\phi$ depends only on the external spacetime coordinates,$\phi = \phi(x^\mu)$. $\alpha$ and $\beta$ are arbitrary numbers which we will choose later in order to compare the results with four-dimensional theory. Using this ansatz, the Ricci scalar can be written as
\begin{eqnarray}
\sqrt{-g}R = \sqrt{-\bar{g}}\sqrt{\tilde{g}} e^{(2\alpha + n\beta)\phi} \Big(&\,& \bar{R} + e^{2(\alpha-\beta)}\tilde{R}-2(3\alpha+n\beta) \Box\phi \nonumber\\
&-& (6\alpha^2 + n(n+1) \beta^2 + 4n\alpha\beta) (\partial\phi)^2\Big).
\label{kkR1}
\end{eqnarray}
We note that the quantities with ``bar", $\bar{X}$, stand for quantities in $(3+1)$-external spacetime and the quantities with ``tilde", $\tilde{X}$, stand for quantities in $n$-internal space. In order to get our usual $(3+1)$-dimensional spacetime, one can set $\beta = - 2\alpha/n$ to get rid of overall factor of $\bar{R}$. Substituting $\beta$ back into equation (\ref{kkR1}), we have
\begin{eqnarray}
\sqrt{-g}R = \sqrt{-\bar{g}}\sqrt{\tilde{g}}  \left( \bar{R} + e^{\frac{2}{n}(n+2)\alpha\phi}\tilde{R}-2\alpha \Box\phi - \frac{2}{n}(n+2)\alpha^2 (\partial\phi)^2\right).
\label{kkR2}
\end{eqnarray}
By performing in the same way as we find the Ricci scalar, Gauss-Bonnet term can be written as
\begin{eqnarray}
\sqrt{-g}\mathcal{G} = \sqrt{-\bar{g}} \sqrt{\tilde{g}} \Big[& &e^{n\beta\phi} \left( \bar{\mathcal{G}} -4 f_1 \bar{G}^{\mu\nu}\partial_\mu \phi \partial_\nu \phi-2 f_2 (\partial\phi)^2 \Box\phi - f_3 (\partial\phi)^2(\partial\phi)^2\right)\,\,\,\,\,\,\,\, \nonumber\\
 &+& e^{(2\alpha+ (n-2)\beta)\phi}\tilde{R}(\bar{R}+f_4(\partial\phi)^2 ) + e^{(4\alpha + (n-4)\beta)\phi}\tilde{\mathcal{G}}\Big],
\label{kkG1}
\end{eqnarray}
where
\begin{eqnarray}
f_1 &=& 2 n \alpha  \beta  + n(n-1) \beta ^2,\\
f_2 &=& 6n \alpha ^2 \beta  +6 n(n-1)\alpha  \beta ^2+ n(n-1) (n-2)\beta ^3,\\
f_3 &=& 8n \alpha ^3 \beta  +4 n(4n-3)\alpha ^2 \beta ^2+8 n(n-1) ^2\alpha  \beta ^3+8 n(n-1) ^2(n-2)\beta ^4,\\
f_4 &=& 6\alpha ^2+6 (n-2) \alpha  \beta  +(n-2)(n-3) \beta ^2.
\label{fG}
\end{eqnarray}
If we choose the parameter as $\beta = - 2\alpha/n$, Gauss-Bonnet term becomes
\begin{eqnarray}
\sqrt{-g}\mathcal{G} = \sqrt{-\bar{g}} \sqrt{\tilde{g}} \Big[& &e^{-2\alpha\phi} \left( \bar{\mathcal{G}} -4 f_1 \bar{G}^{\mu\nu}\partial_\mu \phi \partial_\nu \phi-2 f_2 (\partial\phi)^2 \Box\phi - f_3 (\partial\phi)^2(\partial\phi)^2\right)\,\,\,\,\,\,\,\,\,\,\, \nonumber\\
 &+& e^{4\alpha\phi/n}\tilde{R}(\bar{R}+f_4(\partial\phi)^2 ) + e^{2(n-4)\alpha\phi/n}\tilde{\mathcal{G}}\Big].
\label{kkG2}
\end{eqnarray}
The cosmological constant and matter terms can be obtained in the same way and then written respectively as
\begin{eqnarray}
\sqrt{-g}\lambda^{-2} &=& \sqrt{-\bar{g}} \sqrt{\tilde{g}} \,e^{(4\alpha+ n\beta)\phi}\lambda^{-2} = \sqrt{-\bar{g}} \sqrt{\tilde{g}} \,e^{2\alpha\phi}\lambda^{-2}, \\
\sqrt{-g}L_{m} &=& \sqrt{-\bar{g}} \sqrt{\tilde{g}} \,e^{(4\alpha+ n\beta)\phi}L_{m} = \sqrt{-\bar{g}} \sqrt{\tilde{g}} \,e^{2\alpha\phi}L_{m}.
\label{kkG2}
\end{eqnarray}
We assume that $\tilde{g}_{\mu\nu}$ is a Euclidean metric. Note that we also use this assumption for calculating the Casimir energy density. Hence, we can integrate out the extra dimension coordinates. To obtain the Newton's gravitational constant in four-dimensional spacetime, the mass scale will relate to the Planck mass as follow
\begin{eqnarray}
M^2_{Pl}= M^{n+2}_{*}\int d^n y \sqrt{\tilde{g}}  = M^{n+2}_{*} (2\pi)^n.
\label{M4Mn}
\end{eqnarray}
Since $\tilde{g}_{\mu\nu}$ is a Euclidean metric, the $\tilde{R} = \tilde{\mathcal{G}} = 0$. The EGB action can be written as
\begin{eqnarray}
S_{EGB} = \int d^4 x \sqrt{-\bar{g}}\frac{M^{2}_{Pl}}{2}\Big[&& \lambda^2 e^{-2\alpha\phi} \left( \bar{\mathcal{G}} -4 f_1 \bar{G}^{\mu\nu}\partial_\mu \phi \partial_\nu \phi-2 f_2 (\partial\phi)^2 \Box\phi - f_3 (\partial\phi)^2(\partial\phi)^2\right) \nonumber \\
&-& \frac{2 e^{2\alpha\phi}}{ \lambda^{2}} + \Big( \bar{R} - \frac{2}{n}(n+2)\alpha^2 (\partial\phi)^2\Big) + \frac{2 (2\pi)^n}{M^{2}_{Pl}} e^{2\alpha\phi}L_{m}  \Big],
\end{eqnarray}
where we have set $\alpha_0 = -2, \alpha_1 = \alpha_2 =1$. In order to obtain the canonical form of the scalar field, one has to set
\begin{eqnarray}
\alpha = -\sqrt{\frac{n}{4(n+2)}},
\end{eqnarray}
and then the EGB action becomes
\begin{eqnarray}
S_{EGB} = \int d^4 x \sqrt{-\bar{g}}\frac{M^{2}_{Pl}}{2} \Big[&& \lambda^2 e^{\sqrt{\frac{n}{(n+2)}}\phi} \left( \bar{\mathcal{G}} -4 f_1 \bar{G}^{\mu\nu}\partial_\mu \phi \partial_\nu \phi-2 f_2 (\partial\phi)^2 \Box\phi - f_3 (\partial\phi)^2(\partial\phi)^2\right) \nonumber \\
&-& 2 e^{-\sqrt{\frac{n}{(n+2)}}\phi} \lambda^{-2} + \Big( \bar{R} - \frac{1}{2} (\partial\phi)^2\Big) + \frac{2 (2\pi)^n}{M^{2}_{Pl}} e^{-\sqrt{\frac{n}{(n+2)}}\phi}L_{m}\Big].
\end{eqnarray}
The EGB theory in five-dimensional spacetime will be reduced to
\begin{eqnarray}
S_{EGB} = \int d^4 x \sqrt{-\bar{g}}\Big[& & \frac{M^{2}_{Pl}}{2} \Big\{ - 2\, e^{-\frac{\psi}{\psi_0}} \lambda^{-2} + \bar{R} + \lambda^2 e^{\frac{\psi}{\psi_0}} \bar{\mathcal{G}}  \Big\}+ 2\pi e^{-\frac{\psi}{\psi_0}}L_{m} \nonumber\\
&-& \frac{1}{2} (\partial\psi)^2+ \lambda^2e^{\frac{\psi}{\psi_0}}  \Big(\frac{4}{3} \bar{G}^{\mu\nu}\partial_\mu \psi \partial_\nu \psi+(\partial\psi)^2 \frac{\Box\psi}{\psi_0}\Big)\Big],\label{action-EGB-4}
\end{eqnarray}
where we have rescaled the scalar field as follow
\begin{eqnarray}
\phi=\frac{\sqrt{2}\psi}{M_{Pl}}=\sqrt{3}\frac{\psi}{\psi_0}.
\end{eqnarray}
In order to relate the radius of the extra dimension $b(t)$ with the scalar $\phi(t)$, we note that the explicit relation can be expressed as
\begin{eqnarray}
b(t) = e^{\frac{\phi}{\sqrt{n(n+2)}}}.
\end{eqnarray}
\section{Casimir dark energy model with Einstein-Gauss-Bonnet theory\label{sec:CDE with GB}}
\begin{figure}[htp]
\centering
\includegraphics[width=0.95\textwidth]{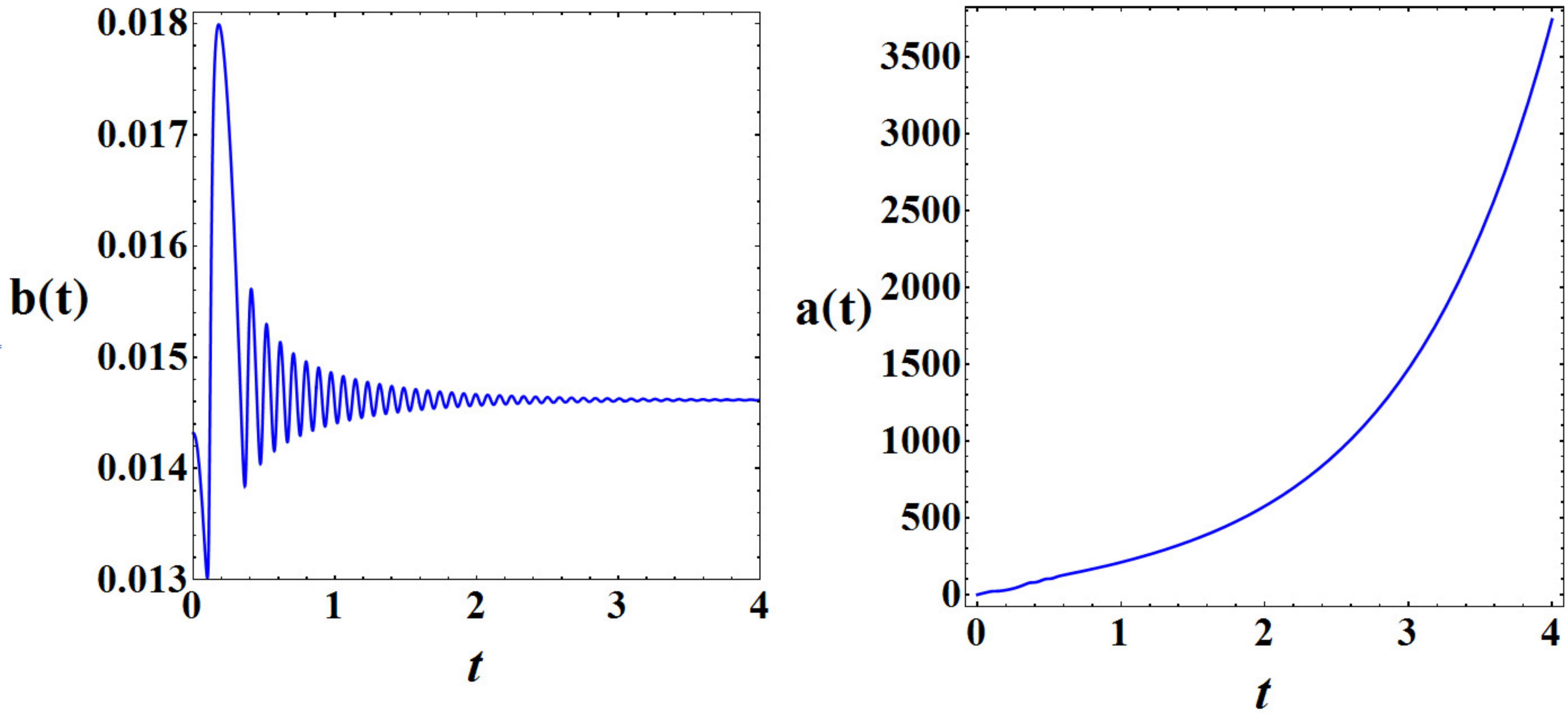}
\caption[The evolution of radius of the extra dimension and the scale factor including non-relativistic matter and Gauss-Bonnet term]{The evolution of radius of the extra dimension (in the left panel) and the scale factor (in the right panel) in the Casimir dark energy model including non-relativistic matter and Gauss-Bonnet term. From the left panel, the radius of the extra dimension can be stabilized and from the right panel our 3-spatial universe can be accelerated. We have used $\lambda = 0.1$ in this simulation. }\label{HbHa}
\end{figure}

In this section, we will consider EGB theory with an action (\ref{D-action-EGB}). We set the zero order of Lovelock invariance as a cosmological constant $\alpha_0 \lambda^{-2}= -2 \Lambda$. In our ansatz, this cosmological constant is interpreted as Casimir energy $ M^{3}_{*}\Lambda = \rho_{Cas} = L_{Cas}(\phi)$. By setting $\alpha_1=\alpha_2 =1$, the action can be expressed as
\begin{equation}
S_{5} = \int d^5 x \sqrt{-g} \left( \frac{M^{3}_{*}}{2} ( R + \lambda^2 \mathcal{G})- L_{Cas}(\phi) + L_m \right).\label{action-EGB-5}
\end{equation}
The equation of motion are obtained from equation (\ref{D-eom-EGB}) as
\begin{equation}
 G^\mu_\nu + \lambda^2 H^\mu_\nu = M^{-3}_{*} \Big(T^\mu_{\nu (Cas)}+T^\mu_{\nu (m)} \Big),\label{eom-EGB-5}
\end{equation}
where $T^\mu_{\nu (Cas)}$ is defined in equation (\ref{Cas emt}). By using the metric in (\ref{flrw}), each components of the above equation can be expressed as
\begin{eqnarray}
3H_a^2+ 3H_a H_b +12 \lambda^2 H_a^3 H_b &=& M^{-3}_{*}(\rho_{Cas} + \rho_m),\label{eom1}\\
\frac{\ddot{b}}{b}+2\frac{\ddot{a}}{a}+H_a^2 +2H_a H_b +\lambda^2 \left(8 H_a H_b \frac{\ddot{a}}{a}+4 H_a^2\frac{\ddot{b}}{b}\right)&=& -M^{-3}_{*}p_a,\label{eom2}\\
3\frac{\ddot{a}}{a}+ 3H_a^2 +\lambda^2  H_a^2 \frac{\ddot{a}}{a} &=& - M^{-3}_{*} p_b. \label{eom3}
\end{eqnarray}
The numerical results of these equations of motion are illustrated in Figure \ref{HbHa}. We can see that the extra dimension can be stabilized and the universe is accelerated. Note that we use $\lambda = 0.1$ in the simulation. The range in which the extra dimension can be stabilized is approximated as $0.005 \leq \lambda^2 \leq 5.0$. This range depends on the initial velocity of the extra dimension, $\dot{b}_i$. By setting $\lambda = 1$, the initial velocity is required as $\dot{b}_i < 0.002$ for stabilizing the extra dimension. To compare the results of the model with the standard history of the universe, we need to include the contribution of radiation. This contribution may alter the behavior of the stabilizing mechanism. This modification can be done by changing the initial conditions to be the value at the radiation-dominated period and substituting the energy density and pressure of the radiation instead of matter. By doing this, we found that stabilizing mechanism still hold. In order to see how the stabilization mechanism can be restored, we will analyze this mechanism in effective equation of motion in four-dimensional spacetime.

\section{Dynamics in radion picture\label{sec:Radion}}
By using the reduced action in equation (\ref{action-EGB-4}) and our interpretation of cosmological constant term, the 4-dimensional spacetime can be written as
\begin{eqnarray}
S^{(4)} = \int d^4 x \sqrt{-\bar{g}}\Big[& & \frac{M^{2}_{Pl}}{2} \Big\{ \bar{R} + \lambda^2 e^{\frac{\psi}{\psi_0}} \bar{\mathcal{G}}  \Big\} - \frac{1}{2} (\partial\psi)^2 - V(\psi)+ e^{-2\frac{\psi}{\psi_0}}L_{m}^{(4)}\nonumber \\
&+& \lambda^2e^{\frac{\psi}{\psi_0}}  \Big(\frac{4}{3} \bar{G}^{\mu\nu}\partial_\mu \psi \partial_\nu \psi+(\partial\psi)^2 \frac{\Box\psi}{\psi_0}\Big)\Big],\label{action-4}
\end{eqnarray}
where $V(\psi) = 2 \pi e^{-\frac{\psi}{\psi_0}}L_{Cas}^{(5)}(\psi) = e^{-2\frac{\psi}{\psi_0}}L_{Cas}^{(4)}(\psi)$. Note that without the last two terms and potential term in this action, it is found that this action can be derived from heterotic or IIB string and there is a tachyonic instability for the tensor mode and tachyon free for the scalar mode \cite{Kawai:1998ab, Kawai:1999pw}. Study of this model also is investigated in Bianchi type I \cite{Kawai:1998bn} and type IX \cite{Kim:2013xu}. It is found that this kind of the model seems to be unstable. However, by including the last two terms, the stability conditions in the tensor mode will be modified. With out the potential term, theory seem to be stable but the matter phase is absent \cite{Gannouji:2011qz, Kobayashi:2011nu}. Adding the potential term corresponding to Casimir energy may provide the matter phase in the history of the universe. We leave this investigation for further work. The components of the equation of motion obtained by varying this action with respect to $g_{\mu\nu}$ can be written as
\begin{eqnarray}
3 M_{Pl}^2 H_a^2 &=& \rho_\psi + e^{-2\frac{\psi}{\psi_0}} \rho_m^{(4)},\label{eom4D-1}\\
M_{Pl}^2 (3H_a^2 + 2 \dot{H_a})&=&- p_\psi,\label{eom4D-2}
\end{eqnarray}
where
\begin{eqnarray}
\rho_\psi &=& \frac{1}{2}\dot{\psi}^2 + V  + \tilde{\lambda}^2 e^{\frac{\psi}{\psi_0}}\Big( -8 H_a^3 \frac{\dot{\psi}}{\psi_0}+ 12 H_a^2 \frac{\dot{\psi}^2}{\psi_0^2} + 6 H_a \frac{\dot{\psi}^3}{\psi_0^3}-\frac{\dot{\psi}^4}{\psi_0^4} \Big),\\
p_\psi &=& \Big(\frac{1}{2}\dot{\psi}^2 - V\Big)  + \tilde{\lambda}^2 e^{\frac{\psi}{\psi_0}}\Big( \frac{16}{3} H_a(H_a^2+\dot{H_a}) \frac{\dot{\psi}}{\psi_0}- \frac{4}{3} (H_a^2+2\dot{H_a})\frac{\dot{\psi}^2}{\psi_0^2} \nonumber \\
&-&\frac{8}{3} H_a \frac{\dot{\psi}^3}{\psi_0^3}-\frac{\dot{\psi}^4}{\psi_0^4}+\frac{2}{3} (4H_a^2 - 8 H_a\frac{\dot{\psi}}{\psi_0} - \frac{\dot{\psi}^2}{\psi_0^2} )\frac{\ddot{\psi}}{\psi_0} \Big).
\end{eqnarray}
The conservation of the energy momentum tensor provides the field equation for the radion field as
\begin{eqnarray}
\ddot{\psi} + 3H_a\dot{\psi} =- V'_{eff} + 2e^{-2\frac{\psi}{\psi_0}}\frac{\rho_m^{(4)}}{\psi_0},\label{eom4D-3}
\end{eqnarray}
where
\begin{eqnarray}
V'_{eff}=V'+\tilde{\lambda}^2 e^{\frac{\psi}{\psi_0}}\Big( &-&\frac{8}{\psi_0} H_a^2( H_a^2 + \dot{H_a}) +8 H_a (3H_a^2 + 2\dot{H_a})\frac{\dot{\psi}}{\psi_0^2}+ (22 H_a^2 + 6\dot{H_a})\frac{\dot{\psi}^2}{\psi_0^3}\nonumber\\
&-&\frac{\dot{\psi}^4}{\psi_0^5} + 4(2H_a^2 + 3 H_a\frac{\dot{\psi}}{\psi_0} -\frac{\dot{\psi}^2}{\psi_0^2} )\frac{\ddot{\psi}}{\psi_0^2}\Big),\label{Vp4D-eff}\\
\tilde{\lambda}^2 = \psi_0^2\lambda^2 = \frac{3}{2}M_{Pl}^2\lambda^2.
\end{eqnarray}

Considering equation (\ref{eom4D-1}) with $\tilde{\lambda} = 0$, neglecting the effect of Gauss-Bonnet term, one can see that the contribution from matter field will dominate at the early time since $\rho^{(4)}_m \propto a^{-3}$ and $a \ll 1$. Therefore, the minimum of the potential will disappear. From equation (\ref{eom4D-3}) and (\ref{Vp4D-eff}) with $\tilde{\lambda} = 0$, the contribution from matter field also increases the slope of the potential. Hence, the radion field will roll down rapidly and then passes away from the minimum point before the minimum of the potential exists. This is the destabilizing mechanism of the extra dimension discussed in \cite{Greene:2007xu}.

Now we will see how Gauss-Bonnet contribution alters the dynamics of the radion field. From equation (\ref{eom4D-1}), the effect of Gauss-Bonnet term does not significantly change the existence of the potential minimum at the early time since the effect of matter field is still dominant. However, the contribution of Gauss-Bonnet term can significantly change the dynamics of the radion field through the slope of the potential as seen in equation (\ref{Vp4D-eff}). Initially, the radion field is put in some points in the potential away from the minimum with a tiny fraction of velocity, $\dot{\psi}/\psi_i \ll 0$. Note that this assumption is also required in order to stabilize the extra dimension in normal Casimir dark energy model. Therefore, the first term from Gauss-Bonnet contribution in equation (\ref{Vp4D-eff}) is dominant and effectively reduces the slope of the potential corresponding to reduce the magnitude of the force acting on the radion field. Note that, at the beginning time, $\ddot{a}/a = \dot{H}_a+H_a^2 < 0$ and $V' < 0$. This term effectively slow down the radion field and eventually the radion field does not pass minimum point before it exists.

Our analysis can also be applied in six-dimensional spacetime. The procedure can be evaluated in the same manner as we have done in five-dimensional spacetime. We show here only the reduced action and the significant changes of the equations of motion.
The reduced action from six-dimensional spacetime can be written as
\begin{eqnarray}
S^{(4)} = \int d^4 x \sqrt{-\bar{g}}\Big[& & \frac{M^{2}_{Pl}}{2} \Big\{ \bar{R} + \lambda^2 e^{\frac{\psi}{\psi_0}} \bar{\mathcal{G}}  \Big\} - \frac{1}{2} (\partial\psi)^2 - V(\psi) \nonumber \\
&+& \lambda^2e^{\frac{\psi}{\psi_0}} \Big( \bar{G}^{\mu\nu}\partial_\mu \psi \partial_\nu \psi - \frac{(\partial\psi)^2(\partial\psi)^2}{4\psi_0^2}\Big)+ e^{-2\frac{\psi}{\psi_0}}L_{m}^{(4)}\Big],\,\,\,\,\,\,\label{action6to4}
\end{eqnarray}
where  $V(\psi) = (2 \pi)^2 e^{-\frac{\psi}{\psi_0}}L_{Cas}^{(6)}(\psi) = e^{-2\frac{\psi}{\psi_0}}L_{Cas}^{(4)}(\psi)$ and the constant $\psi_0$ is now redefined as $\psi_0 = M_{Pl}$. The field $\phi$ and $\psi$ are related together as $\phi = \sqrt{2}\psi / \psi_0 $.
The energy density, pressure and $V'_{eff}$ can be rewritten as
\begin{eqnarray}
\rho_\psi &=& \frac{1}{2}\dot{\psi}^2 + V  + \tilde{\lambda}^2 e^{\frac{\psi}{\psi_0}}\Big( -12 H_a^3 \frac{\dot{\psi}}{\psi_0}+ 9 H_a^2 \frac{\dot{\psi}^2}{\psi_0^2} -\frac{3}{4}\frac{\dot{\psi}^4}{\psi_0^4} \Big),\\
p_\psi &=& \Big(\frac{1}{2}\dot{\psi}^2 - V\Big)  + \tilde{\lambda}^2 e^{\frac{\psi}{\psi_0}}\Big( 8 H_a(H_a^2+\dot{H_a}) \frac{\dot{\psi}}{\psi_0}+ (H_a^2-2\dot{H_a})\frac{\dot{\psi}^2}{\psi_0^2} \nonumber \\
&-&2 H_a \frac{\dot{\psi}^3}{4\psi_0^3}-\frac{\dot{\psi}^4}{\psi_0^4}+4 (H_a^2 - 2H_a\frac{\dot{\psi}}{\psi_0} )\frac{\ddot{\psi}}{\psi_0} \Big),\,\,\,\,\,\\
V'_{eff} &=& V'+\tilde{\lambda}^2 e^{\frac{\psi}{\psi_0}}\Big(-\frac{12}{\psi_0} H_a^2( H_a^2 + \dot{H_a}) +6 H_a (3H_a^2 + 2\dot{H_a})\frac{\dot{\psi}}{\psi_0^2}\nonumber \\
&-&3 H_a^2 \frac{\dot{\psi}^2}{\psi_0^3} -\frac{3}{4}\frac{\dot{\psi}^4}{\psi_0^5} + 3(2H_a^2 -\frac{\dot{\psi}^2}{\psi_0^2} )\frac{\ddot{\psi}}{\psi_0^2}\Big).\label{Vp6to4D-eff}
\end{eqnarray}
At the beginning time, $\dot{\psi}/\psi_i \ll 0$, then $V'_{eff} $ can be approximated as
\begin{eqnarray}
V'_{eff} \sim V'-\tilde{\lambda}^2 e^{\frac{\psi}{\psi_0}}\frac{12}{\psi_0} H_a^2( H_a^2 + \dot{H_a}).\label{Vp6to4D-eff-app}
\end{eqnarray}
Again, the effective force acting on the radion field is reduce at the beginning time. Using the same manner in five-dimensional spacetime, the decreasing of the effective force will lead to restoring of the stabilization mechanism of the extra dimensions. This analysis is confirmed by using the numerical simulation of the equations of motion in six-dimensional spacetime. In six-dimensional spacetime, we have 9 degrees of freedom for the graviton. We choose the number of degrees of freedom to be the number of degrees of freedom for the massless boson in the total Casimir energy density. The other numbers are obtained in such a way that the potential must has a local minimum. Our ansatz in six-dimensional spacetime can be written as
\begin{equation}
\rho_{Cas}^{(6)} = 9 \rho_{boson}^{massless} + 14 \rho_{fermion}^{massless} + 14 \rho_{boson}^{massive} + 14 \rho_{fermion}^{massive},
\label{sum_casimir 6D}
\end{equation}
where the mass ratio is $\bar{\lambda} = 0.534$. The range of the parameter $\lambda$ can be obtained as $0.3 < \lambda < 2.0$. The range is sensitive to the mass ratio and also depends on the initial value of $b(t)$. These parameters may also be constrained from the observations, for example \cite{Amendola:2005cr,Koivisto:2006xf,Amendola:2007ni}. The constraint of the parameters to the observation is out of scope of this work. We leave it for the further work. It is important to note that the allow region of the parameters may not be consistent with the stability condition in order to avoid ghost degrees of freedom in the model \cite{Stelle:1977ry,Barth:1983hb, DeFelice:2006pg, Calcagni:2006ye}. We also note that, by replacing the non-relativistic matter with the radiation, the stabilizing mechanism in six-dimensional spacetime still hold.

\section{Conclusions}\label{sec:conclusions}
The concept of the Casimir dark energy model is reviewed. The important idea of this model is that it is natural to interpret the Casimir energy emerging from compactification of the extra dimensions as dark energy to drive the late-time accelerating universe \cite{Greene:2007xu}. However, this model of dark energy encounters the problem in which the extra dimensions cannot be stabilized when the non-relativistic matter is taken into account. One of the solutions of this problem is that adding the exotic field such as aether field into the model \cite{Chatrabhuti:2009ew}. However, the aether theory by itself is not stable \cite{Carroll:2009em,Himmetoglu:2008hx,Himmetoglu:2008zp,Himmetoglu:2009qi}. In this paper, we seek for another solution by generalizing the Einstein gravity theory to Einstein-Gauss-Bonnet (EGB) gravity theory. It is worthy to investigate EGB theory since it is a generalization of Einstein gravity theory  in higher-dimensional spacetime by keeping second order derivative of the equation of motion and satisfying the conservation equation of matter field. It is also compatible with the low-energy effective field theory of string theory \cite{Antoniadis:1993jc,Campbell:1990fu}. The result of our investigation in five-dimensional spacetime shows that the radius of extra dimension can be stabilized when the Gauss-Bonnet (GB) term and non-relativistic matter are taken into account. The extension by including radiation into the model is also investigated and the result shows that the stabilizing mechanism still hold. We use the radion picture in four-dimensional spacetime to analyze how GB term can provide the stabilization mechanism. It is found that the Gauss-Bonnet contribution effectively reduces the slope of the radion potential at the beginning time corresponding to reduce the magnitude of the force acting on the radion field. Therefore, the radion field slowly rolls down and does not pass minimum point of the effective potential before the minimum the potential exists and then the stabilizing mechanism is restored eventually. We also investigate this behavior in six-dimensional spacetime. The extra dimensions can also be stabilized in the same manner as we analyze in five-dimensional spacetime. The stability of the model by itself is the important issue for investigating. The effective four-dimensional GB theory with exponential potential and without terms corresponding to nonminimal coupling to the gravity are found to be unstable due to tachyonic instability \cite{Calcagni:2006ye, Tsujikawa:2006ph}. The investigation of the model by including the nonminimal coupling terms without potential term implies that the model is stable but the matter phase is absent in the history of the universe \cite{Gannouji:2011qz, Kobayashi:2011nu}. Adding the potential term corresponding to the Casimir energy may provide this matter phase. We leave this investigation including the constraints of the theoretical parameters for further work. The interplay between the Gauss-Bonnet term and the dynamical radion field in our model may shed some light on the connection between the modified gravity theory and the fundamental high-energy theories which have a requirement of higher-dimensional spacetime.

\begin{acknowledgments}
The author would like to thank String Theory and Supergravity Group, Department of Physics, Faculty of Science, Chulalongkorn University for hospitality during this work was in progress. He is also grateful to Parinya Karndumri and Antonio De Felice for helpful conversations and comments in the manuscript. This work is supported by Naresuan University Research Fund through grant R2556C042.
\end{acknowledgments}

\end{document}